\documentclass[a4paper,12pt,english]{article}

\usepackage{amsfonts,bm,amssymb,euscript,array,babel,cite}
\usepackage{amsmath,amsthm} 
\usepackage[dvips]{epsfig}



\newcommand{\be}{\begin{equation}} \newcommand{\ee}{\end{equation}}
\newcommand{\beq}{\begin{equation}} \newcommand{\eeq}{\end{equation}}
\newcommand{\beqa}{\begin{eqnarray}}
\newcommand{\eeqa}{\end{eqnarray}} \newcommand{\eq}[1]{(\ref{#1})}
\def\nn{\nonumber} \def\bea{\begin{eqnarray}} \def\eea{\end{eqnarray}}
\def\obar{\overline}
\def\a{\alpha}  
  
 \def\d{\delta}

   \def\cD{{\cal
D}}   
  \def\cJ{{\cal J}}    \def\cN{{\cal N}}

\def\R{{\mathbb R}}  \def\N{{\mathbb N}}
\def\Z{{\mathbb Z}} \def\one{\mbox{1 \kern-.59em {\rm l}}}

\def\mmu{\mathfrak{u}}
\def\msu{\mathfrak{su}}

\def\bit{\begin{itemize}} \def\eit{\end{itemize}} \def\Tr{\mbox{Tr}}

\def\({\left(} \def\){\right)}    

\renewcommand{\Tr}{{\rm Tr}}

 \allowdisplaybreaks[3]

\begin{document}

\renewcommand{\title}[1]{\vspace{10mm}\noindent{\Large{\bf
#1}}\vspace{8mm}} \newcommand{\authors}[1]{\noindent{\large
#1}\vspace{5mm}} \newcommand{\address}[1]{{\itshape #1\vspace{2mm}}}

\begin{titlepage}
\begin{flushright}
UWThPh-2006-12\\
Dista-UPO/06\\
\end{flushright}

\begin{center}

\title{ \Large Dynamical generation of fuzzy extra dimensions, \\[1ex]
dimensional reduction and symmetry breaking }

\vskip 3mm

\authors{Paolo {\sc Aschieri${}^1$}, Theodoros {\sc
Grammatikopoulos${}^2$}, \\[1ex] Harold {\sc Steinacker${}^3$}, George
{\sc Zoupanos${}^2$}}

\vskip 3mm

\address{ ${}^1$ Dipartimento di Scienze e Tecnologie Avanzate,\\
Universit{\'a} del Piemonte Orientale, and INFN,\\ Corso Borsalino 54,
I-15100, Alessandria, Italy\\[1ex] ${}^2$ Physics Department National
Technical University\\ Zografou Campus, GR-15780 Athens\\[1ex]
${}^3$Institut f\"ur Theoretische Physik, Universit\"at Wien\\
Boltzmanngasse 5, A-1090 Wien, Austria\\[3ex] E-mail:
{aschieri@theorie.physik.uni-muenchen.de,\\ tgrammat@mail.ntua.gr,\\
harold.steinacker@univie.ac.at,\\ George.Zoupanos@cern.ch}}

\vskip 1.4cm

\textbf{Abstract}

\vskip 3mm

\begin{minipage}{14cm}%

We present a renormalizable 4-dimensional $SU(\cN)$ gauge theory with
a suitable multiplet of scalar fields, which dynamically develops 
extra dimensions in the form of  a fuzzy sphere $S^2_N$. We explicitly find
the tower of massive Kaluza-Klein modes consistent with an
interpretation as gauge theory on $M^4 \times S^2$, the scalars 
being interpreted as gauge fields on $S^2$.  
The gauge group is broken dynamically, and the low-energy
content of the model is determined. Depending on the parameters of the
model the low-energy gauge group can be $SU(n)$, or broken further to
$SU(n_1) \times SU(n_2) \times U(1)$, with mass scale determined by
the size of the extra dimension.

\end{minipage}

\end{center}

\end{titlepage}

 \tableofcontents

\section{Introduction}

It is difficult to overestimate the relevance of the Kaluza-Klein
programme of unification in higher dimensions. 
In this beautiful programme, higher dimensions are an input however, and the
4-dimensional theory has to be recovered. We here reverse the logic and
see how a 4-dimensional gauge theory dynamically develops 
higher dimensions. 
The very concept of dimension
therefore gets an extra, richer
dynamical perspective. For pioneering work in that context 
see \cite{Arkani-Hamed:2001ca}. 
Furthermore, the Kaluza-Klein programme can now 
be pursued within the
framework of a 4-dimensional field theory, which dynamically develops 
higher dimensions.

We present in this paper a simple field-theoretical model 
which realizes that idea.
It is defined as a renormalizable $SU(\cN)$ gauge
theory on 4-dimensional Minkowski space $M^4$, containing 3 scalars in the
adjoint of $SU(\cN)$ that transform as vectors under an additional global
$SO(3)$ symmetry with the most general renormalizable potential.  We
then show that the model dynamically develops fuzzy extra dimensions,
more precisely a fuzzy sphere $S^2_{N}$. The appropriate
interpretation is therefore as gauge theory on $M^4 \times S^2_{N}$. The
low-energy effective action is that of a 4-dimensional gauge theory on
$M^4$, whose gauge group and field content is dynamically determined
by compactification and dimensional reduction 
on the internal sphere  $S^2_{N}$.  An
interesting and quite rich pattern of spontaneous symmetry breaking appears,
breaking the original $SU(\cN)$ gauge symmetry down to much smaller
and potentially quite interesting low-energy gauge groups.  In
particular, we find explicitly the tower of massive Kaluza-Klein
states, which justifies the interpretation as a compactified
higher-dimensional gauge theory. Nevertheless, the model is
renormalizable.

A different mechanism of dynamically generating extra dimensions has
been proposed some years ago in \cite{Arkani-Hamed:2001ca}, known
under the name of ``deconstruction''.  In this context, renormalizable
4-dimensional asymptotically free gauge theories were considered with
suitable Moose- or Quiver-type arrays of gauge groups and couplings,
which develop a ``lattice-like'' fifth dimension. This idea attracted
considerable interest.  Our model is quite different, and very simple:
The $SU(\cN)$ gauge theory with 3 scalars $\phi_a$ in the
adjoint and a global $SO(3)$ symmetry is shown to develop fuzzy extra
dimensions through a symmetry breaking mechanism.

Let us discuss some of the features of our model in more detail.  The
effective geometry, the symmetry breaking pattern and the low-energy
gauge group are determined dynamically in terms of a few free
parameters of the potential. We discuss in detail the two simplest
possible vacua
with gauge groups $SU(n)$ and $SU(n_1)\times SU(n_2) \times U(1)$. We
find explicitly the tower of massive Kaluza-Klein modes corresponding
to the effective geometry.  The mass scale of these massive gauge
bosons is determined by the size of the extra dimensions, which in
turn depends on some logarithmically running coupling constants.  In
the case of the $SU(n_1)\times SU(n_2) \times U(1)$ vacuum, we
identify in particular massive gauge fields in the bifundamental,
similar as in GUT models with an adjoint Higgs.  Moreover, we also
identify a candidate for a further symmetry breaking mechanism, which
may lead to a low-energy content of the theory close to the standard
model.

There is no problem in principle to add fermions to our model.  In
particular, we point out that in the vacua with low-energy gauge
group $SU(n_1) \times SU(n_2)\times U(1)$, the extra-dimensional
sphere always carries a magnetic flux with nonzero monopole
number. This is very interesting in the context of fermions,
since internal fluxes naturally lead to chiral massless fermions.
However, this is a delicate issue and will be discussed in a
forthcoming paper.

Perhaps the most remarkable aspect of our model is that the geometric
interpretation and the corresponding low-energy degrees of freedom
depend in a nontrivial way on the parameters of the model, which are
running under the RG group. 
Therefore the massless degrees of freedom and their geometrical 
interpretation depend on the energy scale. 
In particular, the low-energy gauge group generically turns out to be
$SU(n_1) \times SU(n_2)\times U(1)$ or $SU(n)$, while
 gauge groups which are
products of more than two simple components (apart from $U(1)$) do not
seem to occur in this model. Moreover, the values of $n_1$ and $n_2$ 
are determined dynamically, and may well be small such as 3 and 2. 
A full analysis of the hierarchy of all possible vacua and their
symmetry breaking pattern is not trivial however, and will not be
attempted in this paper. Here we restrict ourselves to establish the
basic mechanisms and features of the model, and discuss in section 3
the two simplest cases (that we name ``type 1'' and ``type 2'' vacuum) 
in some detail. A more detailed
analysis (in particular for the ``type 3 vacuum'') is left for future
work.

The idea to use fuzzy spaces for the extra dimensions is certainly 
not new.
This work was motivated by a fuzzy coset space dimensional reduction
(CSDR) scheme considered recently in
\cite{Aschieri:2003vy,Aschieri:2005wm,Aschieri:2004vh}, combined with
lessons from the matrix-model approach to gauge theory on the fuzzy
sphere \cite{Steinacker:2003sd,Steinacker:2004yu}. This leads in
particular to a dynamical mechanism of determining the vacuum, SSB
patterns and background fluxes.  A somewhat similar model has been
studied recently in \cite{Andrews:2005cv,Andrews:2006aw}, which
realizes deconstruction and a ``twisted'' compactification of an extra
fuzzy sphere based on a supersymmetric gauge theory. Our model is
different and does not require supersymmetry, leading to a much richer
pattern of symmetry breaking and effective geometry.  
For other relevant work see e.g. \cite{Madore:1992ej}.

The dynamical
formation of fuzzy spaces found here is also related to recent 
work studying the emergence of stable submanifolds 
in modified IIB matrix models. 
In particular, previous studies based on  actions for
fuzzy gauge theory different from ours generically only 
gave results corresponding to
$U(1)$ or $U(\infty)$ gauge groups, see e.g.
\cite{Azuma:2004ie,Azuma:2005bj,Azuma:2004zq}
and references therein. 
The dynamical generation of a nontrivial index on noncommutative 
spaces has also been observed in \cite{Aoki:2004sd,Aoki:2006zi} for  
different models.

Our mechanism may also be very interesting in the context of the
recent observation \cite{Abel:2005rh} that extra dimensions are very 
desirable for the application of noncommutative field theory to 
particle physics. Other related recent work discussing the 
implications of the higher-dimensional point of view on symmetry
breaking and Higgs masses can be found in
\cite{Lim:2006bx,Dvali:2001qr,Antoniadis:2002ns,Scrucca:2003ra}. These
issues could now be discussed within a renormalizable framework.

\section{The 4-dimensional action}
 
We start with a $SU(\cN)$ gauge theory on 4-dimensional Minkowski
space $M^4$ with coordinates $y^\mu$, $\mu = 0,1,2,3$.  The action
under consideration is 
\be {\cal S}_{YM}= \int d^{4}y\, Tr\,\left(
\frac{1}{4g^{2}}\, F_{\mu \nu}^\dagger F_{\mu \nu} +
(D_{\mu}\phi_{{a}})^\dagger D_{\mu}\phi_{{a}}\right) - V(\phi)
\label{the4daction}
\ee 
where $A_\mu$ are $\msu(\cN)$-valued gauge fields, $D_\mu =
\partial_\mu + [A_\mu,.]$, and 
\be \phi_{{a}} = - \phi_{{a}}^\dagger
~, \qquad a=1,2,3 
\ee 
are 3  antihermitian scalars in the
adjoint of $SU(\cN)$, \be \phi_{{a}} \to U^\dagger \phi_{{a}} U \ee
where $U = U(y) \in SU(\cN)$. Furthermore, the $\phi_a$ transform as
vectors of an additional global $SO(3)$ symmetry. The potential
$V(\phi)$ is taken to be the most general renormalizable action
invariant under the above symmetries, which is
\begin{eqnarray}
V(\phi) &=& Tr\, \left( g_1 \phi_a\phi_a \phi_b\phi_b +
g_2\phi_a\phi_b\phi_a \phi_b - g_3 \varepsilon_{a b c} \phi_a \phi_b
\phi_c + g_4\phi_a \phi_a \right) \nn\\ && + \frac{g_5}{\cN}\,
Tr(\phi_a \phi_a)Tr(\phi_b \phi_b) + \frac{g_6}{\cN} Tr(\phi_a
\phi_b)Tr(\phi_a \phi_b) +g_7.
\label{pot}
\end{eqnarray}
This may not look very transparent at first sight, however it can be
written in a very intuitive way. First, we make the scalars
dimensionless by rescaling 
\be \phi'_a = R\; \phi_a, 
\ee 
where $R$ has
dimension of length; we will usually suppress $R$ since it can
immediately be reinserted, and drop the prime from now on.  Now
observe that for a suitable choice of $R$, 
\be
R = \frac{2 g_2}{g_3}, 
\label{Radius}
\ee 
the potential can be rewritten as \bea V(\phi)&=& Tr \( a^2
(\phi_a\phi_a + \tilde b\, \one)^2 + c +\frac 1{\tilde g^2}\,
F_{ab}^\dagger F_{ab}\,\) + \frac{h}{\cN}\, g_{ab} g_{ab}
\label{V-general-2}
\eea for suitable constants $a,b,c,\tilde g,h$, where \bea F_{{a}{b}}
&=& [\phi_{{a}}, \phi_{{b}}] - \varepsilon_{abc} \phi_{{c}}\, =
\varepsilon_{abc} F_c , \nn\\ \tilde b &=& b + \frac{d}{\cN} \,
Tr(\phi_a \phi_a), \nn\\ g_{ab} &=& Tr(\phi_a \phi_b).
\label{const-def}
\eea We will omit $c$ from now.  The potential is clearly positive
definite provided 
\be a^2 = g_1+g_2 >0, \qquad \frac 2{\tilde g^2} = -
g_2 >0, \qquad h \geq 0, 
\ee 
which we assume from now on.  Here $\tilde b =
\tilde b(y)$ is a scalar, $g_{ab} = g_{ab}(y)$ is a symmetric
tensor under the global $SO(3)$, and $F_{ab}=F_{ab}(y)$ is a
$\msu(\cN)$-valued antisymmetric tensor field which will be
interpreted as field strength in some dynamically generated extra
dimensions below.  In this form, $V(\phi)$ looks like the action of
Yang-Mills gauge theory on a fuzzy sphere in the matrix formulation
\cite{Steinacker:2003sd,Steinacker:2004yu,Carow-Watamura:1998jn,
Presnajder:2003ak}.  The presence of the first term
$a^2 (\phi_a\phi_a + \tilde b)^2$ might seem strange at first, however
we should not simply omit it since it would be reintroduced by
renormalization. In fact it is necessary for the interpretation as YM
action, 
and we will see that it is very welcome on physical grounds since it
dynamically determines and stabilizes a vacuum, which can be
interpreted as extra-dimensional fuzzy sphere. In particular, it
removes unwanted flat directions.

Let us briefly comment on the RG flow of the various
constants. Without attempting any precise computations here, we can
see by looking at the potential \eq{pot} that $g_4$ will be
quadratically divergent at one loop, while $g_1$ and $g_2$ are
logarithmically divergent. Moreover, the only diagrams contributing to
the coefficients $g_5, g_6$ of the ``nonlocal'' terms are nonplanar,
and thus logarithmically divergent but suppressed by $\frac 1\cN$
compared to the other (planar) diagrams. This justifies the explicit
factors $\frac 1\cN$ in \eq{pot} and \eq{const-def}.  Finally, the
only one-loop diagram contributing to $g_3$ is also logarithmically
divergent. In terms of the constants in the potential
\eq{V-general-2}, this implies that $R, a$, $\tilde g$, $d$ and $h$
are running logarithmically under the RG flux, while $b$ and therefore
$\tilde b$ is running quadratically.
The gauge coupling $g$ is of course logarithmically divergent and 
asymptotically free.

A full analysis of the RG flow of these parameters is complicated by
the fact that the vacuum and the number of massive resp. massless
degrees of freedom depends sensitively on the values of these
parameters, as will be discussed below. This indicates that the RG
flow of this model will have a rich and nontrivial structure, with
different effective description at different energy scales.

\subsection{The minimum of the potential}
\label{sec:vacua}

Let us try to determine the minimum of the potential
\eq{V-general-2}. This turns out to be a rather nontrivial task, and
the answer depends crucially on the parameters in the potential.

For suitable values of the parameters in the potential, we can
immediately write down the vacuum. 
Assume  for simplicity $h=0$ in \eq{V-general-2} . Since
$V(\phi) \geq 0$, the global minimum of the potential is certainly
achieved if 
\be F_{{a}{b}} = [\phi_{{a}}, \phi_{{b}}] -
\varepsilon_{abc} \phi_{{c}}~ = 0, \qquad -\phi_a\phi_a = \tilde b,
\label{vacuum-trivial-cond}
\ee 
because then $V(\phi) =0$. This implies that $\phi_a$ is a
representation of $SU(2)$, with prescribed Casimir\footnote{note that
$-\phi\cdot \phi = \phi^\dagger\cdot \phi >0$ since the fields are
antihermitian} $\tilde b$. These equations may or may not have a
solution, depending on the value of $\tilde b$.  Assume first that
$\tilde b$ coincides with the quadratic Casimir of a
finite-dimensional irrep of $SU(2)$, 
\be \tilde b = C_2(N) = \frac 14
(N^2-1) 
\ee for some $N \in \N$. If furthermore the dimension $\cN$ of
the matrices $\phi_a$ can be written as \be \cN = N n, \ee then
clearly the solution of \eq{vacuum-trivial-cond} is given by 
\be
\phi_a = X_a^{(N)} \otimes \one_{n}
\label{vacuum-trivial}
\ee 
up to a gauge transformation, where $X_a^{(N)}$ denote the
generator of the $N$-dimensional irrep of $SU(2)$. This can be
viewed as a special case of \eq{solution-general} below, consisting of
$n$ copies of the irrep $(N)$ of $SU(2)$.

For generic $\tilde b$, the equations \eq{vacuum-trivial-cond} cannot
be satisfied for finite-dimensional matrices $\phi_a$. The exact
vacuum (which certainly exists since the potential is positive
definite) can in principle be found by solving the ``vacuum equation''
$\frac{\d V}{\d \phi_a} =0$, 
\be a^2 \{\phi_a,\phi\cdot\phi + \tilde b
+ \frac{d}{\cN}\, Tr(\phi\cdot \phi+\tilde b)\} + \frac{2h}{\cN}
g_{ab}\phi_b + \frac 1{\tilde g^2}\, (2[F_{ab},\phi_b] + F_{bc}
\varepsilon_{abc}) =0
\label{eom-int}
\ee 
where $\phi\cdot\phi = \phi_a \phi_a$. We note that all solutions
under consideration will imply $g_{ab} = \frac 13 \delta_{ab}
Tr(\phi\cdot \phi)$, simplifying this expression.

The general solution of \eq{eom-int} is not known. However, it is easy
to write down a large class of solutions: any decomposition of $\cN =
n_1 N_1 + ... + n_h N_h$ into irreps of $SU(2)$ with multiplicities
$n_i$ leads to a block-diagonal solution 
\be \phi_a = diag\Big(\a_1\,
X_a^{(N_1)}, ..., \a_k\, X_a^{(N_k)}\Big)
\label{solution-general}
\ee 
of the vacuum equations \eq{eom-int}, where $\a_i$ are suitable
constants which will be determined below. There are hence several
possibilities for the true vacuum, i.e. the global minimum of the
potential. Since the general solution is not known, we proceed by
first determining the solution of the form \eq{solution-general} with
minimal potential, and then discuss a possible solution of a different
type (``type 3 vacuum'').

\paragraph{Type 1 vacuum.}

It is clear that the solution with minimal potential should satisfy
\eq{vacuum-trivial-cond} at least approximately. It is therefore
plausible that the solution \eq{solution-general} with minimal
potential contains only representations whose Casimirs are close to
$\tilde b$. In particular, let $N$ be the dimension of the irrep whose
Casimir $C_2(N)\approx \tilde b$ is closest to $\tilde b$. If
furthermore the dimensions match as $\cN = N n$, we expect that the
vacuum is given by $n$ copies of the irrep $(N)$, which can be written
as \be \phi_a = \a\, X_a^{(N)} \otimes\one_{n}.
\label{vacuum-mod1}
\ee This is a slight generalization of \eq{vacuum-trivial}, with $\a$
being determined through the vacuum equations \eq{eom-int}, \be a^2
(\a^2 C_2(N) - \tilde b)(1+d) + \frac h3 \a^2 C_2(N) - \frac 1{\tilde
g^2}\, (\a-1)(1-2\a) =0
\label{eom-alpha}
\ee 
A vacuum of the form \eq{vacuum-mod1} will be denoted as ``type 1
vacuum''.  As we will explain in detail, it has a natural
interpretation in terms of a dynamically generated extra-dimensional
fuzzy sphere $S^2_{N}$, by interpreting $X_a^{(N)}$ as generator of a
fuzzy sphere \eq{fuzzycoords}. Furthermore, we will show in section
\ref{sec:KK1} that this type 1 vacuum \eq{vacuum-mod1} leads to
spontaneous symmetry breaking, with low-energy (unbroken) gauge group
$SU(n)$. The low-energy sector of the model can then be understood as
compactification and dimensional reduction on this internal fuzzy
sphere.

Let us discuss equation \eq{eom-alpha} in more detail. It can of
course be solved exactly, but an expansion around $\a=1$ is more
illuminating.  To simplify the analysis we assume 
\be 
d=h=0 
\ee from
now on, and assume furthermore that 
\be 
a^2 \approx \frac 1{\tilde g^2} 
\ee have the same order of magnitude.  Defining the {\em real}
number $\tilde N$ by 
\be \tilde b = \frac 14 (\tilde N^2-1), \ee one
finds 
\be \a = 1 -\frac {m}{N} + \frac{m(m+1)}{N^2} + O(\frac 1{N^3})
\qquad \mbox{where} \,\, m = N - \tilde N,
\label{alpha-solution-N}
\ee 
assuming $N$ to be large and $m$ small. Notice that $a$ does not enter to
leading order. This can be understood by noting that the first term in
\eq{eom-alpha} is dominating under these assumptions, which determines
$\a$ to be \eq{alpha-solution-N} to leading order. The potential
$V(\phi)$ is then dominated by the term 
\be \frac {1}{\tilde g^2}
F_{ab}^\dagger F_{ab} = \frac {1}{2\tilde g^2}\, m^2 \,\one\,\, + \,
O(\frac 1{N}),
\label{action-m}
\ee 
while $(\phi_a\phi_a + \tilde b)^2 = O(\frac 1{N^2})$.  There is a
deeper reason for this simple result: If $\,\tilde N \in \N$, then the
solution \eq{vacuum-mod1} can be interpreted as a fuzzy sphere
$S^2_{\tilde N}$ carrying a magnetic monopole of strength $m$, as
shown explicitly in \cite{Steinacker:2003sd}; see also
\cite{Karabali:2001te,Balachandran:1999hx}. Then \eq{action-m} is
indeed the action of the monopole field strength.

\paragraph{Type 2 vacuum.} 
It is now easy to see that for suitable parameters, the vacuum will
indeed consist of several distinct blocks. This will typically be the
case if $\cN$ is not divisible by the dimension of the irrep whose
Casimir is closest to $\tilde b$.

Consider again a solution \eq{solution-general} with $n_i$ blocks of
size $N_i = \tilde N +m_i$, assuming that $\tilde N$ is large and
$\frac{m_i}{\tilde N} \ll 1$.  Generalizing \eq{action-m}, the action
is then given by 
\be V(\phi) = Tr \Big( \frac {1}{2\tilde g^2}\,
\sum_i n_i\, m_i^2 \, \one_{N_i} \, + O(\frac 1{N_i}) \Big) \approx
\frac{1}{2\tilde{g}^{2}}\, \frac{\cN}{k}\, \sum_{i} n_i \, m_{i}^{2}\,
\label{action-mi}
\ee 
where $k=\sum n_i$ is the total number of irreps, and the solution
can be interpreted in terms of ``instantons'' (nonabelian monopoles)
on the internal fuzzy sphere \cite{Steinacker:2003sd}.  Hence in order
to determine the solution of type \eq{solution-general} with minimal
action, we simply have to minimize $\sum_i n_i \, m_i^2$, where the
$m_i \in \Z -\tilde N$ satisfy the constraint $\sum n_i \,m_i = \cN -
k \tilde N$.

It is now easy to see that as long as the approximations used in
\eq{action-mi} are valid, the vacuum is given by a partition 
consisting of  blocks with no more than 2
distinct sizes $N_1, N_2$ which satisfy $N_2 = N_1+1$.
The follows from the convexity of \eq{action-mi}: 
assume that the vacuum is given by a 
configuration with 3 or more different blocks of size 
$N_1 < N_2 < ... < N_k$. Then the action \eq{action-mi}
could be lowered by modifying the configuration as follows:
reduce $n_1$ and $n_k$ by one, and add 2 blocks of size
$N_1+1$ and $N_k-1$. This preserves the overall dimension, and 
it is easy to check (using convexity) 
that the action \eq{action-mi} becomes smaller. 
This argument can be applied as long as 
there are 3 or more different blocks, or 2 blocks with 
$|N_2 - N_1| \geq 2$. Therefore if
$\cN$ is large, the solution with minimal potential among all possible
partitions \eq{solution-general} is given either by a type 1 vacuum, or
takes the form 
\be \phi_a
= \left(\begin{array}{cc}\a_1\, X_a^{(N_1)}\otimes\one_{n_1} & 0 \\ 0
& \a_2\,X_a^{(N_2)}\otimes\one_{n_2}
             \end{array}\right),
\label{vacuum-mod2}
\ee where the integers $N_1, N_2$ satisfy 
\be 
\cN = N_1 n_1 + N_2 n_2,
\qquad N_2 = N_1+1.
\ee
A vacuum of the form \eq{vacuum-mod2} will be denoted as ``type 2
vacuum'', and is the generic case.
In particular, the integers $n_1$ and $n_2$ are determined dynamically.
This conclusion might be altered for nonzero $d,h$ or by a violation
of the approximations used in \eq{action-mi}.  We will show in section
\ref{sec:KK2} that this type of vacuum leads to a low-energy
(unbroken) gauge group $SU(n_1) \times SU(n_2) \times U(1)$, and the
low-energy sector can be interpreted as dimensional reduction of a
higher-dimensional gauge theory on an internal fuzzy sphere, with
features similar to a GUT model with SSB $SU(n_1+n_2) \to SU(n_1)
\times SU(n_2) \times U(1)$ via an adjoint Higgs. Furthermore, since
the vacuum \eq{vacuum-mod2} can be interpreted as a fuzzy sphere with
nontrivial magnetic flux \cite{Steinacker:2003sd}, one can expect to
obtain massless chiral fermions in the low-energy action. This will be
worked out in detail in a forthcoming publication.

In particular, it is interesting to see that gauge groups which are
products of more than two simple components (apart from $U(1)$) do not
occur in this model.

\paragraph{Type 3 vacuum.}
Finally, it could be that the vacuum is of a type different from
\eq{solution-general}, e.g. with off-diagonal corrections such as \be
\phi_a = \left(\begin{array}{cc}\a_1\, X_a^{(N_1)}\otimes\one_{n_1} &
\varphi_a \\ - \varphi_a^\dagger & \a_2\,X_a^{(N_2)}\otimes\one_{n_2}
             \end{array}\right)
\label{vacuum-mod3}
\ee for some small $\varphi_a$.  We will indeed provide evidence for
the existence of such a vacuum below, and argue that it leads to a
further SSB.  This might play a role similar to low-energy
(``electroweak'') symmetry breaking, which will be discussed in more
detail below. In particular, it is interesting to note that the
$\varphi_a$ will no longer be in the adjoint of the low-energy gauge
group.  A possible way to obtain a SSB scenario close to the standard
model is discussed in section \ref{sec:standardmodel}.

\subsection{Emergence of extra dimensions and the fuzzy sphere}
\label{sec:emergence}

Before discussing these vacua and the corresponding symmetry breaking
in more detail, we want to explain the geometrical interpretation,
assuming first that the vacuum has the form \eq{vacuum-mod1}. The
$X_a^{(N)}$ are then interpreted as coordinate functions (generators)
of a fuzzy sphere $S^2_{N}$, and the ``scalar'' action 
\be S_{\phi} =
Tr V(\phi) = Tr\Big(a^2 (\phi_a\phi_a + \tilde b)^2 + \frac 1{\tilde
g^2}\, F_{ab}^\dagger F_{ab}\Big)
\label{S-YM2}
\ee 
for $\cN \times \cN$ matrices $\phi_a$ is precisely the action for
a $U(n)$ Yang-Mills theory on $S^2_{N}$ with coupling $\tilde g$, as
shown in \cite{Steinacker:2003sd} and reviewed in
section \ref{sec:fuzzygaugetheory}. In fact, the ``unusual'' term
$(\phi_a\phi_a + \tilde b)^2$ is essential for this interpretation,
since it stabilizes the vacuum $\phi_a = X_a^{(N)}$ and gives a large
mass to the extra ``radial'' scalar field which otherwise arises.  The
fluctuations of $\phi_a = X_a^{(N)} + A_a$ then provide the components
$A_a$ of a higher-dimensional gauge field $A_M = (A_\mu, A_a)$, and
the action \eq{the4daction} can be interpreted as YM theory on the
6-dimensional space $M^4 \times S^2_{N}$, with gauge group depending
on the particular vacuum.  Note that e.g. for the type 1 vacuum, the
local gauge transformations $U(\cN)$ can indeed be interpreted as
local $U(n)$ gauge transformations on $M^4 \times S^2_{N}$.

In other words, the scalar degrees of freedom $\phi_a$ conspire to
form a fuzzy space in extra dimensions.  We therefore interpret the
vacuum \eq{vacuum-mod1} as describing dynamically generated extra
dimensions in the form of a fuzzy sphere $S^2_{N}$, with an induced
Yang-Mills action on $S^2_{N}$. This geometrical interpretation will
be fully justified in section \ref{sec:KK} by working out the spectrum
of Kaluza-Klein modes.  The effective low-energy theory is then given
by the zero modes on $S^2_{N}$, which is analogous to the models
considered in \cite{Aschieri:2003vy}. However, in the present approach
we have a clear dynamical selection of the geometry due to the first
term in \eq{S-YM2}.

It is interesting to recall here the running of the coupling constants
under the RG as discussed above. The logarithmic running of $R$
implies that the scale of the internal spheres is only mildly affected
by the RG flow. However, $\tilde b$ is running essentially
quadratically, hence is generically large. This is quite welcome here:
starting with some large $\cN$, $\tilde b \approx C_2(\tilde N)$ must
indeed be large in order to lead to the geometric interpretation
discussed above. Hence the problems of naturalness or fine-tuning 
appear to be rather mild here.

\section{Kaluza-Klein modes,
dimensional reduction, and symmetry breaking}
\label{sec:KK}

We now study the model \eq{the4daction} in more detail.  Let us
emphasize again that this is a 4-dimensional renormalizable gauge
theory, and there is no fuzzy sphere or any other extra-dimensional
structure to start with. We have already discussed possible vacua of
the potential \eq{S-YM2}, depending on the parameters $a,\tilde
b,\tilde g$ and $\cN$. This is a nontrivial problem, the full solution
of which is beyond the scope of this paper. We restrict ourselves here
to the simplest types of vacua discussed in section \ref{sec:vacua},
and derive some of the properties of the resulting low-energy models,
such as the corresponding low-energy gauge groups and the excitation
spectrum.  In particular, we exhibit the tower of Kaluza-Klein modes
in the different cases. This turns out to be consistent with an
interpretation in terms of compactification on an internal sphere,
demonstrating without a doubt the emergence of fuzzy internal
dimensions.  In particular, the scalar fields $\phi_a$ become
gauge fields on the fuzzy sphere.

\subsection{Type 1 vacuum and $SU(n)$ gauge group}
\label{sec:KK1}

Let us start with the simplest case, assuming that the vacuum has the
form \eq{vacuum-mod1}. We want to determine the spectrum and the
representation content of the gauge field $A_\mu$.  The structure of
$\phi_a = \a\, X_a^{(N)} \otimes\one_{n}$ suggests to consider the
subgroups $SU(N) \times SU(n) $ of $SU(\cN)$, where \be K:=SU(n) \ee
is the commutant of $\phi_a$ i.e. the maximal subgroup of $SU(\cN)$
which commutes with all $\phi_a$, $a=1,2,3$; this follows from Schur's
Lemma. $K$ will turn out to be the effective (low-energy) unbroken
4-dimensional gauge group.

We could now proceed in a standard way arguing that $SU(\cN)$ is
spontaneously broken to $K$ since $\phi_a$ takes a VEV as in
\eq{vacuum-mod1}, and elaborate the Higgs mechanism. This is
essentially what will be done below, however in a language which is
very close to the picture of compactification and KK modes on a sphere
in extra dimensions. This is appropriate here, and leads to a
description of the low-energy physics of this model as a dimensionally
reduced $SU(n)$ gauge theory.

\paragraph{Kaluza-Klein expansion on $S^2_{N}$.}

Interpreting the $X_a^{(N)}$ as generators of the fuzzy sphere
$S^2_{N}$, we can decompose the full 4-dimensional $\msu(\cN)$-valued
gauge fields $A_\mu$ into spherical harmonics $Y^{lm}(x)$ on
the fuzzy sphere $S^2_{N}$ with coordinates $x_a$:
\be 
A_\mu = \sum_{0 \leq l\leq N,
|m|\leq l} \, Y^{lm}(x)\otimes A_{\mu,lm}(y) = A_\mu(x,y).
\label{KK-modes-A}
\ee 
The $Y^{lm}$ are by definition irreps under
the $SU(2)$ rotations on $S^2_{N}$, and form a basis of Hermitian
$N \times N$ matrices; for more details see section \ref{sec:fuzzysphere}.  
The $A_{\mu,lm}(y)$ turn out to be $\mmu(n)$-valued
gauge and vector fields on $M^4$. Using this expansion, we can
interpret $A_\mu(x,y)$ as $\mmu(n)$-valued functions on $M^4\times
S^2_{N}$, expanded into the Kaluza-Klein modes (i.e. harmonics) of
$S^2_{N}$.

The scalar fields $\phi_a$ with potential \eq{S-YM2} and vacuum
\eq{vacuum-mod1} should be interpreted as ``covariant coordinates'' on
$S^2_{N}$ which describe $U(n)$ Yang-Mills theory on $S^2_{N}$. This
means that the fluctuations $A_a$ of these covariant coordinates 
\be
\phi_a = \a\, X_a^{(N)} \otimes\one_{n} + A_a 
\ee 
should be
interpreted as gauge fields on the fuzzy sphere, see
\eq{gaugefield-S2N}. They can be expanded similarly as 
\be A_a =
\sum_{l,m} \, Y^{lm}(x)\otimes A_{a,lm}(y) = A_a(x,y),
\label{KK-modes-phi}
\ee interpreted as functions (or 1-form) on $M^4\times S^2_{N}$ taking
values in $\mmu(n)$. One can then interpret $A_M(x,y) = (A_\mu(x,y),
A_a(x,y))$ as $\mmu(n)$-valued gauge or vector fields on $M^4\times
S^2_{N}$.

Given this expansion into KK modes, we will show that only
$A_{\mu,00}(y)$ (i.e. the dimensionally reduced gauge field) becomes a
massless $\msu(n)$-valued\footnote{note that $A_{\mu,00}(y)$ is
traceless, while $A_{\mu,lm}(y)$ is not in general} gauge field in 4D,
while all other modes $A_{\mu,lm}(y)$ with $l \geq 1$ constitute a
tower of Kaluza-Klein modes with large mass gap, and decouple for low
energies. The existence of these KK modes firmly establishes our claim
that the model develops dynamically extra dimensions in the form of
$S^2_{N}$.
This geometric interpretation is hence forced upon us, provided the
vacuum has the form \eq{vacuum-mod1}.  The scalar fields $A_a(x,y)$
will be analyzed in a similar way below, and provide no additional
massless degrees of freedom in 4 dimensions. More complicated vacua
will have a similar interpretation.  Remarkably, our model is fully
renormalizable in spite of its higher-dimensional character, in
contrast to the commutative case; see also \cite{Aschieri:2005wm}.

\paragraph{Computation of the KK masses.}

To justify these claims, let us compute the masses of the KK modes
\eq{KK-modes-A}. They are induced by the covariant derivatives $\int
Tr (D_{\mu}\phi_{{a}})^{2} $ in \eq{the4daction}, 
\be \int Tr
(D_{\mu}\phi_{{a}})^\dagger D_{\mu}\phi_{{a}} = \int Tr
(\partial_{\mu}\phi_{{a}}^\dagger \partial_{\mu}\phi_{{a}} + 2
(\partial_{\mu}\phi_{{a}}^\dagger) [A_\mu,\phi_{a}] +
[A_\mu,\phi_{a}]^\dagger[A_\mu,\phi_{a}]).
\label{covar-mass-term}
\ee 
The most general scalar field configuration can be written as 
\be
\phi_a(y) = \a(y) X_a^{(N)} \otimes \one_n + A_a(x,y)
\label{phi-a-expansion}
\ee 
where $A_a(x,y)$ is interpreted as gauge field on the fuzzy sphere
$S^2_{N}$ for each $y \in M^4$. We allow here for a $y$--dependent
$\a(y)$ (which could have been absorbed in $A_a(x,y)$), because it is
naturally interpreted as the Higgs field responsible for the symmetry
breaking $SU(\cN) \to SU(n)$. As usual, the last term in
\eq{covar-mass-term} leads to the mass terms for the gauge fields
$A_\mu$ in the vacuum $\phi_a(y) = \a X_a^{(N)} \otimes \one_n$,
provided the mixed term which is linear in $A_\mu$ vanishes in a
suitable gauge. This is usually achieved by going to the unitary
gauge. In the present case this is complicated by the fact that we
have 3 scalars in the adjoint, and there is no obvious definition of
the unitary gauge; in fact, there are are too many scalar degrees of
freedom as to gauge away that term completely. However, we can choose
a gauge where all quadratic contributions of that term vanish, leaving
only cubic interaction terms. To see this, we insert
\eq{phi-a-expansion} into the term $(\partial_{\mu}\phi_{{a}}^\dagger)
[A_\mu,\phi_{a}]$ in \eq{covar-mass-term}, which gives
$$ 
\int Tr A_\mu [\phi_a,\partial_{\mu}\phi_{{a}}^\dagger] = \int Tr
A_\mu \Big(\a [X_a,\partial_{\mu} A_a(x,y)] +
[A_a(x,y),\partial_{\mu}\a\, X_a] +
[A_a(x,y),\partial_{\mu}A_a(x,y)]\Big).
$$ 
Now we partially fix the gauge by imposing the ``internal'' Lorentz
gauge $[X_a, A_a] =0$ at each point $y$. This is always
possible\footnote{even though this gauge is commonly used in the
literature on the fuzzy sphere, a proof of existence has apparently
not been given.  It can be proved by extremizing the real function
$Tr(X_a \phi_a)$ on a given gauge orbit, which is compact; the
e.o.m. then implies $[X_a, \phi_a] =0$.}, and the above simplifies as
\be 
\int Tr A_\mu [\phi_a,\partial_{\mu}\phi_{{a}}^\dagger] = \int Tr
A_\mu [A_a(x,y),\partial_{\mu}A_a(x,y)] =: S_{int}.
\label{A-lin-term-2}
\ee 
This contains only cubic interaction terms, which are irrelevant
for the computation of the masses. We can therefore proceed by setting
$\phi_a(y) = \a X_a^{(N)} \otimes \one_n$ and inserting the expansion
\eq{KK-modes-A} of $A_\mu$ into the last term of
\eq{covar-mass-term}. Noting that $i[X_{a},A_\mu] = J_a A_\mu =
\sum_{l,m} A_{\mu,lm}(y)\,J_{a} Y^{lm}$ is simply the action of
$SU(2)$ on the fuzzy sphere, it follows that $Tr [X_{a},A_\mu]
[X_{a},A_\mu]$ is the quadratic Casimir on the modes of $A_\mu$ which
are orthogonal, and we obtain 
\be \int Tr (D_{\mu}\phi_{{a}})^\dagger
D_{\mu}\phi_{{a}}= \int Tr (\partial_{\mu}\phi_{{a}}^\dagger
\partial_{\mu}\phi_{{a}} + \sum_{l,m} \a^2\, l(l+1)\,
A_{\mu,lm}(y)^\dagger A_{\mu,lm}(y) ) + S_{int}.
\label{covar-mass-term-2}
\ee 
Therefore the 4-dimensional $\mmu(n)$ gauge fields $A_{\mu,lm}(y)$
acquire a mass \be m^2_{l} = \frac{\a^2 g^2}{R^2}\, l(l+1)
\label{KK-masses-Adiag}
\ee reinserting the parameter $R$ \eq{Radius} which has dimension
length. This is as expected for higher KK modes, and determines the
radius of the internal $S^2$ to be 
\be 
r_{S^2} = \frac{\a}g R 
\label{S-radius}
\ee
where $\a\approx 1$ according to \eq{alpha-solution-N}. In particular, only $A_{\mu}(y)\equiv
A_{\mu,00}(y)$ survives as a massless 4-dimensional $\msu(n)$ gauge
field. The low-energy effective action for the gauge sector is then
given by 
\be S_{LEA} = \int d^4 y\,\frac 1{4g^{2}}\, Tr_n \ F_{\mu
\nu}^\dagger F_{\mu \nu},
\label{LEA-action-FF}
\ee 
where $F_{\mu \nu}$ is the field strength of the low-energy
$\msu(n)$ gauge fields, dropping all other KK modes whose mass scale
is set by $\frac 1{R}$.  For $n=1$, there is no massless gauge field. 
However we would find a massless $U(1)$ gauge field if we 
start with a $U(\cN)$ gauge theory rather than  $SU(\cN)$.

\paragraph{Scalar sector.}

We now expand the most general scalar fields $\phi_a$ into modes,
singling out the coefficient of the ``radial mode'' as 
\be \phi_a(y) =
X_a^{(N)} \otimes (\a \one_n + \varphi(y)) + \sum_{k} A_{a,k}(x)
\otimes\varphi_{k}(y).
\label{phi-a-expansion-modes}
\ee 
Here $A_{a,k}(x)$ stands for a suitable basis labeled by $k$ of
fluctuation modes of gauge fields on $S^2_{N}$, and $\varphi(y)$
resp. $\varphi_k(y)$ are $\mmu(n)$-valued.  We expect that all
fluctuation modes in the expansion \eq{phi-a-expansion-modes} have a
large mass gap of the order of the KK scale, which is indeed the case
as shown in detail in section \ref{sec:stability}.  Therefore we can
drop all these modes for the low-energy sector. However, the field
$\varphi(y)$ plays a somewhat special role.  It corresponds to
fluctuations of the radius of the internal fuzzy sphere, which is the
order parameter responsible for the SSB  $SU(\cN) \to SU(n)$, 
and assumes the value $\a \one_n$ in 
\eq{phi-a-expansion-modes}.  $\varphi(y)$ is therefore 
the Higgs which acquires a positive mass term in the broken
phase, which can be obtained by inserting $\phi_a(y) = X_a^{(N)}
\otimes (\a \one_n + \varphi(y))$ into $V(\phi)$. This mass is
dominated by the first term in \eq{V-general-2} (assuming $a^2 \approx
\frac 1{\tilde g^2}$), of order 
\be 
V(\varphi(y)) \approx N\,
\Big(a^2 C_{2}(N)^2 \varphi(y)^{2} + O(\varphi^3)\Big)
\label{Higgs-potential}
\ee 
for large $\cN$ and $N$.
The full potential for $\varphi$ is of course quartic.

We conclude that our model indeed behaves like a $U(n)$ gauge theory
on $M^4\times S^2_{N}$, with the expected tower of KK modes on the
fuzzy sphere 
$S^2_{N}$ of radius \eq{S-radius}. 
The low-energy
effective action is given by  the lowest KK mode, which is
\be 
S_{LEA} = \int d^4 y\, Tr_n \( \frac
1{4g^{2}}\, F_{\mu \nu}^\dagger F_{\mu \nu} + D_{\mu}\varphi(y)
D_{\mu}\varphi(y) \, N C_2(N) + N a^2 C_{2}(N)^2
\varphi(y)^{2} \) + S_{int}
\label{LEA-action}
\ee 
for the $SU(n)$ gauge field $A_{\mu}(y)\equiv A_{\mu,00}(y)$. In 
\eq{LEA-action} we also
keep the Higgs field $\varphi(y)$, even though it acquires a large
mass
\be
m_\varphi^2 = \frac{a^2}{R^2} \, C_2(N)
\ee
reinserting $R$.

\subsection{Type 2 vacuum and 
$SU(n_1)\times SU(n_2)\times U(1)$ gauge group}
\label{sec:KK2}

For different parameters in the potential, we can obtain a different
vacuum, with different low-energy gauge group.  Assume now that the
vacuum has the form \eq{vacuum-mod2}.  The structure of $\phi_a$
suggests to consider the subgroups $(SU(N_1) \times SU(n_1)) \times
(SU(N_2) \times SU(n_2)) \times U(1)$ of $SU(\cN)$, where 
\be
K:=SU(n_1)\times SU(n_2)\times U(1) 
\ee 
is the maximal subgroup of
$SU(\cN)$ which commutes with all $\phi_a$, $a=1,2,3$ (this follows
from Schur's Lemma).  Here the $U(1)$ factor is embedded as 
\be
\mmu(1) \sim\left(\begin{array}{cc} \frac 1{N_1 n_1}\,\one_{N_1\times
n_1} & \\ & -\frac 1{N_2 n_2}\,\one_{N_2\times n_2} \end{array}\right)
\label{U1-embed}
\ee 
which is traceless.  $K$ will again be the effective (low-energy)
4-dimensional gauge group.

We now repeat the above analysis of the KK modes and their effective
4-dimensional mass.  First, we write \be A_{\mu} =
\left(\begin{array}{cc} A_{\mu}^1 & A_{\mu}^+\\ A_{\mu}^- & A_{\mu}^2
\end{array}\right) \ee according to \eq{vacuum-mod2}, where
$(A_{\mu}^+)^\dagger = -A_{\mu}^-$. The masses of the gauge bosons are
again induced by the last term in \eq{covar-mass-term}.  Consider the
term $[\phi_{a},A_\mu] = [\a_1 X^{(N_1)}_a + \a_2 X^{(N_2)}_a,A_\mu]$.
For the diagonal fluctuations $A_{\mu}^{1,2}$, this is simply the
adjoint action of $X^{(N_1)}_a$. For the off-diagonal modes
$A_{\mu}^\pm$, we can get some insight by assuming first $\a_1 =
\a_2$. Then the above commutator is $X^{(N_1)} A_{\mu}^+ - A_{\mu}^+
X^{(N_2)}$, reflecting the representation content $A_{\mu}^+ \in (N_1)
\otimes (N_2)$ and $A_{\mu}^- \in (N_2) \otimes (N_1)$. Assuming
$N_1-N_2 =k>0$, this implies in particular that there are {\em no zero
modes for the off-diagonal blocks}, rather the lowest angular momentum
is $k$. They can be interpreted as being sections on a monopole bundle
with charge $k$ on $S^2_{N_1}$, cf. \cite{Steinacker:2003sd}.  The
case $\a_{1}\neq \a_2$ requires a more careful analysis as indicated
below.  In any case, we can again expand $A_{\mu}$ into harmonics, 
\be
A_\mu = \sum_{l,m} \left(\begin{array}{cc} Y^{lm (N_1)} \,
A_{\mu,lm}^1(y)& Y^{lm (+)}\, A_{\mu,lm}^+(y)\\ Y^{lm(-)}\,
A_{\mu,lm}^-(y) & Y^{lm (N_2)} \, A_{\mu,lm}^2(y) \end{array}\right) =
A_\mu(x,y)
\label{KK-modes-A2}
\ee setting $Y^{lm (N)}= 0$ if $l > 2N$. Then the
$A_{\mu,lm}^{1,2}(y)$ are $\mmu(n_1)$ resp. $\mmu(n_2)$-valued gauge
resp. vector fields on $M^4$, while $A_{\mu,lm}^{\pm}(y)$ are vector
fields on $M^4$ which transform in the bifundamental
$(n_1,\obar{n}_2)$ resp. $(n_2,\obar{n}_1)$ of $\mmu(n_1)\times
\mmu(n_2)$.

Now we can compute the masses of these fields.  For the diagonal
blocks this is the same as in section \ref{sec:KK1}, while the
off-diagonal components can be handled by writing 
\be
Tr([\phi_{a},A_\mu][\phi_{a},A_\mu]) = 2 Tr(\phi_{a} A_\mu\phi_{a}
A_\mu - \phi_{a}\phi_{a} A_\mu A_\mu) .  
\ee This gives 
\bea \int\!\! 
Tr (D_{\mu}\phi_{{a}})^\dagger D_{\mu}\phi_{{a}} \!\! &=& \!\!  \int
\!\!Tr \Big(\partial_{\mu}\phi_{{a}}^\dagger \partial_{\mu}\phi_{{a}}
+ \sum_{l\geq 0} ( m^2_{l,1}\, A_{\mu,lm}^{1\dagger}(y)
A_{\mu,lm}^1(y) + m^2_{l,2}\, A_{\mu,lm}^{2 \dagger}(y)
A_{\mu,lm}^2(y)) \nn\\ && \quad + \sum_{l\geq k} 2 m^2_{l;\pm}
(A_{\mu,lm}^+(y))^\dagger A_{\mu,lm}^+(y)\Big)
\label{covar-mass-term-break}
\eea 
similar as in \eq{covar-mass-term-2}, with the same gauge choice
and omitting cubic interaction terms. In particular, the diagonal
modes acquire a KK mass 
\be m^2_{l,i} = \frac{\a_i^2 g^2}{R^2}\,l(l+1)
\label{KK-masses-Adiag-12}
\ee 
completely analogous to \eq{KK-masses-Adiag}, while the
off-diagonal modes acquire a mass \bea m^2_{l;\pm} &=&
\frac{g^2}{R^2}\,\(\a_1 \a_2\, l(l+1) +
(\a_1-\a_2)(X_2^2\a_2-X_1^2\a_1)\) \nn\\
&\approx & \frac{g^2}{R^2}\, \( l(l+1) + \frac 14 (m_2-m_1)^2 \,\, +
O(\frac 1\cN) \)
\label{mass-offdiag}
\eea 
using \eq{alpha-solution-N} for $\a_i \approx 1$. In particular,
all masses are positive.

We conclude that the gauge fields $A_{\mu,lm}^{1,2}(y)$ have massless
components $A_{\mu,00}^{1,2}(y)$ which take values in $\msu(n_i)$ due
to the KK-mode $l=0$ (as long as $n_i>1$), while the bifundamental
fields $A_{\mu,lm}^{\pm}(y)$ have no massless components. Note that
the mass scales of the diagonal modes \eq{KK-masses-Adiag-12} and the
off-diagonal modes \eq{mass-offdiag} are essentially the same.  This
result is similar to the breaking $SU(n_1+n_2) \to SU(n_1) \times
SU(n_2)\times U(1)$ through an adjoint Higgs, such as in the $SU(5)
\to SU(3)\times SU(2)\times U(1)$ GUT model.  In that case, one also
obtains massive (``ultraheavy'') gauge fields in the bifundamental,
whose mass should therefore be identified in our scenario with the
mass \eq{mass-offdiag} of the off-diagonal massive KK modes
$A_{\mu,lm}^{\pm}(y)$.  The $U(1)$ factor \eq{U1-embed} corresponds to
the massless components $A_{\mu,00}^{1,2}(y)$ above, which is now
present even if $n_i=1$.
We therefore found results comparable to \cite{Horvath:1977st}, but within the
framework of a renormalizable theory.

The appropriate interpretation of this vacuum is as a gauge theory on
$M^4 \times S^2$, compactified on $S^2$ which carries a magnetic flux
with monopole number $|N_1-N_2|$. This leads to a low-energy action
with gauge group $SU(n_1) \times SU(n_2)\times U(1)$. The existence of
a magnetic flux is particularly interesting in the context of
fermions, since internal fluxes naturally lead to chiral massless
fermions. This issue will be studied in detail elsewhere.

Repeating the analysis of fluctuations for the scalar fields is
somewhat messy, and will not be given here. However since the vacuum
\eq{vacuum-mod2} is assumed to be stable, all fluctuations in the
$\phi_a$ will again be massive with mass presumably given by the KK
scale, and can therefore be omitted for the low-energy theory.  Again,
one could interpret the fluctuations $\varphi_{1,2}(y)$ of the radial
modes $X_a^{(N_{1,2})} \otimes (\a_{1,2} +\varphi_{1,2}(y))$ as
low-energy Higgs in analogy to \eq{phi-a-expansion-modes}, responsible
for the symmetry breaking $SU(n_1+n_2) \to SU(n_1) \times SU(n_2)
\times U(1)$.

 \subsection{Type 3 vacuum and further symmetry breaking}
\label{sec:KK3}

 Finally consider a vacuum of the form \eq{vacuum-mod3}. The
 additional fields $\varphi_{a}$ transform in the bifundamental of
 $SU(n_1) \times SU(n_2)$ and lead to further SSB.  Of particular
 interest is the simplest case \be \phi_a =
 \left(\begin{array}{cc}\a_1\, X_a^{(N_1)}\otimes\one_{n} & \varphi_a
 \\ - \varphi_a^\dagger & \a_2\, X_a^{(N_2)}
              \end{array}\right)
 \label{vacuum-mod4}
 \ee corresponding to a would-be gauge group $SU(n) \times U(1)$
 according to section \ref{sec:KK2}, which will be broken further.
 Then $\varphi_a = \(\begin{array}{c}\varphi_{a,1} \\ \vdots \\
 \varphi_{a,n}\end{array} \)$ lives in the fundamental of $SU(n)$
 charged under $U(1)$, and transforms as $(N_1) \otimes (N_2)$ under
 the $SO(3)$ corresponding to the fuzzy sphere(s). As discussed below,
 by adding a further block, one can get somewhat close to the standard
 model, with $\varphi_a$ being a candidate for a low-energy Higgs.

 We will argue that there is indeed such a solution of the equation of
 motion \eq{eom-int} for $|N_1-N_2|=2$. Note that since $\varphi_a \in
 (N_1) \otimes (N_2) = (|N_1-N_2|+1) \oplus ... \oplus (N_1+N_2-1)$,
 it can transform as a vector under $SO(3)$ only in that case. Hence
 assume $N_1=N_2+2$, and define $\varphi_a \in (N_1) \otimes (N_2)$ to
 be the unique component which transform as a vector in the adjoint.
 One can then show that 
\be \phi_a \phi_a = -
 \left(\begin{array}{cc}\a_1^2\, C_2(N_1)\otimes\one_{n_1} -
 \frac{h}{N_1}\, & 0 \\ 0 & \a_2^2\, C_2(N_2) - \frac{h}{N_2}\,
 \end{array}\right)
 \label{phiphi-2}
 \ee 
where $h$ is a normalization constant, and \bea \varepsilon_{abc}
 \phi_b \phi_c &=& \left(\begin{array}{cc} (\a_1^2- \frac{g_1}{N_1 }\,
 \frac{h}{C_2(N_1)})\,X_a^{(N_1)} & (\a_1 g_1 + \a_2 g_2) \varphi_a \\
 -(\a_1 g_1 + \a_2 g_2) \varphi_a^\dagger & (\a_2^2- \frac{g_2}{N_2
 }\,\frac{h}{C_2(N_2)})\,X_a^{(N_2)} \end{array}\right)
 \label{phiphi-comm-2}
 \eea 
where $g_1=\frac{N_1+1}2, \, g_2=-\frac{N_2-1}2$.  This has the
 same form as \eq{vacuum-mod4} but with different parameters. We now
 have 3 parameters $\a_1,\a_2,h$ at our disposal, hence generically
 this Ansatz will provide solutions of the e.o.m. \eq{eom-int} which
 amounts to 3 equations for the independent blocks.  It remains to be
 seen whether they are energetically favorable.  This will be studied
 in a future publication.

\paragraph{The commutant  $K$ and further symmetry breaking.} 

To determine the low-energy gauge group i.e. the maximal subgroup $K$
commuting with the solution $\phi_a$ of type \eq{vacuum-mod4},
consider 
\bea \varepsilon_{abc} \phi_b \phi_c &-& (\a_1 g_1 + \a_2
g_2) \phi_a = \nn\\ && \!\!\!\!\!\!  \!\!\!\!\!\!\!\!\!\!\!\!
\left(\begin{array}{cc} (\a_1^2 - \a_1(\a_1 g_1 + \a_2 g_2)
-\frac{g_1}{N_1 }\, \frac{h}{C_2(N_1)})\,X_a^{(N_1)} & 0 \\ 0 &
\!\!\!\!\!\!\!\!\!\!\!\!  (\a_2^2 - \a_2(\a_1 g_1 + \a_2 g_2) -
\frac{g_2}{N_2 }\,\frac{h}{C_2(N_2)})\,X_a^{(N_2)}
    \end{array}\right)  \nn\\
\label{phiphi-comm-4}
\eea 
Unless one of the two coefficients vanishes, this implies that
$K$ must commute with \eq{phiphi-comm-4}, hence $K =
\(\begin{array}{cc} K_1 & 0\\0 & K_2\end{array}\)$ is a subgroup of
$SU(n_1) \times SU(n_2)\times U(1)$; here we focus on $SU(n_2) =
SU(1)$ being trivial. Then \eq{vacuum-mod4} implies that $k_1
\varphi_a = \varphi_a k_2$ for $k_i \in K_i$, which means that
$\varphi_a$ is an eigenvector of $k_1$ with eigenvalue $k_2$. Using a
$SU(n_1)$ rotation, we can assume that $\varphi_a^T =
(\varphi_{a,1},0, \dots, 0)$. Taking into account the requirement that
$K$ is traceless, it follows that $K \cong K_1 \cong SU(n_1-1) \subset
SU(n_1)$.  Therefore the gauge symmetry is broken to $SU(n_1-1)$. This
can be modified by adding a further block as discussed below.

\subsection{Towards the standard model}
\label{sec:standardmodel}

Generalizing the above considerations, we can construct a vacuum which
is quite close to the standard model. Consider \be \cN = N_1 n_1 + N_2
n_2 + N_3, \ee for $n_1=3$ and $n_2=2$. As discussed above, we expect
a vacuum of the form \be \phi_a = \left(\begin{array}{ccc} \a_1\,
X_a^{(N_1)}\otimes\one_{3}& 0 &0\\ 0 &
\a_2\,X_a^{(N_2)}\otimes\one_{2} & \varphi_a \\ 0 & -\varphi_a^\dagger
& \a_3\,X_a^{(N_3)}
             \end{array}\right)
\label{vacuum-mod5}
\ee if $\tilde b \approx C_2(N_1)$ and $N_1 \approx N_2 = N_3\pm 2$.
Then the unbroken low-energy gauge group would be \be K = SU(3) \times
U(1)_Q \times U(1)_F, \ee with $U(1)_F$ generated by the traceless
generator 
\be u(1)_F \sim\left(\begin{array}{cc} \frac 1{3
N_1}\,\one_{3 N_1} & \\ & -\frac 1{D}\,\one_{D} \end{array}\right) 
\ee
where $D = 2 N_2 + N_3$, and $U(1)_Q$ generated by the traceless
generator 
\be 
u(1)_Q \sim\left(\begin{array}{ccc} \frac 1{3
N_1}\,\one_{3 N_1} & & \\ & -\frac 1{N_2} \left(\begin{array}{cc} 0 &
0\\ 0 & 1 \end{array}\right) \one_{N_2} & \\ & & 0 \end{array}\right).
\ee 
assuming that $\varphi_a^T = (\varphi_{a,1}, 0)$.  This is
starting to be reminiscent of the standard model, and will be studied
in greater detail elsewhere. However, we should recall that the
existence of a {\em vacuum} of this form has not been established at
this point.

\subsection*{Relation with CSDR scheme}

Let us compare the results of this paper with the CSDR construction in
\cite{Aschieri:2003vy}. In that paper, effective 4-dimensional models
are constructed starting from gauge theory on $M^4 \times S^2_\cN$, by
imposing 
CSDR constraints following the general ideas of 
\cite{Forgacs:1979zs,Kapetanakis:1992hf,Kubyshin:1989vd,Bais:1985yd}. 
These constraints
boiled down to choosing embeddings $\,\omega_a$, $a=1,2,3$ of $SU(2)
\subset SU(\cN)$, which determine the unbroken gauge field 
as the commutant of $\,\omega_a$,
and the low-energy (unbroken) Higgs by
$\varphi_a \sim \omega_a$. This is similar to the ``choice'' of vacuum in
the present paper, such as \eq{vacuum-mod1}, \eq{vacuum-mod2},
identifying $\omega_a$ with $\oplus_i X_a^{N_i}$ as in
\eq{solution-general}.  The solutions of these constraints can be
formally identified with the zero modes $A_{\mu,00}$ of the KK-tower
of gauge fields \eq{KK-modes-A}, resp. the vacuum of the scalar sector
\eq{phi-a-expansion-modes}.  In this sense, the possible vacua
\eq{solution-general} could be interpreted as solutions of the CSDR
constraints in \cite{Aschieri:2003vy} on a given fuzzy sphere.

However, there are important differences. First, the present
approach provides a clear dynamical mechanism which chooses a unique
vacuum. This depends crucially on the first term in \eq{V-general-2}, 
that removes the degeneracy of 
all possible embeddings of $SU(2)$, which have 
vanishing field strength $F_{ab}$.  Moreover, it may provide an
additional mechanism for further symmetry breaking as discussed in
section \ref{sec:KK3}.  Another difference is that the starting point
in \cite{Aschieri:2003vy} is a 6-dimensional gauge theory with some
given gauge group, such as $U(1)$. This is  not the case in present
paper, where the 6-dimensional gauge
group depends on the parameters of the model. 

\section{Discussion}

We have presented a renormalizable 4-dimensional $SU(\cN)$ gauge
theory with a suitable multiplet of scalars, which dynamically develops
fuzzy extra dimensions that form a fuzzy sphere. The model can
then be interpreted as 6-dimensional gauge theory, with gauge group
and geometry depending on the parameters in the original Lagrangian.
We explicitly find the tower of massive Kaluza-Klein modes, consistent
with an interpretation as compactified higher-dimensional gauge
theory, and determine the effective compactified gauge theory.
Depending on the parameters of the model the low-energy gauge group
can be $SU(n)$, or broken further e.g. to $SU(n_1) \times SU(n_2)
\times U(1)$, with mass scale determined by the extra dimension.

There are many remarkable aspects of this model.  First, it provides
an extremely simple and geometrical mechanism of dynamically
generating extra dimensions, without relying on subtle dynamics such
as fermion condensation and particular Moose- or Quiver-type arrays of
gauge groups and couplings, such as in \cite{Arkani-Hamed:2001ca} and
following work. Rather, our model is based on a basic lesson from
noncommutative gauge theory, namely that noncommutative or fuzzy
spaces can be obtained as solutions of matrix models. The mechanism is
quite generic, and does not require fine-tuning or supersymmetry.
This provides in particular a realization of the basic ideas of
compactification and dimensional reduction within the framework of
renormalizable quantum field theory. Moreover, we are essentially 
considering a large $\cN$ gauge theory, which should allow
to apply the analytical techniques developed in this context. 

One of the main features of our mechanism is that the effective
properties of the model including its geometry depend on the
particular parameters of the Lagrangian, which are subject to
renormalization. In particular, the  RG flow of these parameters 
depends on the specific vacuum i.e. geometry,
which in turn will depend on the energy scale.  
For example, it could
be that the model assumes a ``type 3'' vacuum as discussed in section
\ref{sec:KK3} at low energies, which might be quite close to the
standard model.
At higher energies, the parameter $\tilde b$ (which determines the
effective gauge group and which is expected to run quadratically under
the RG flow) will change, implying a very different vacuum with
different gauge group etc. This suggests a rich and complicated
dynamical hierarchy of symmetry breaking, which remains to be
elaborated.  

In particular, we have shown that the low-energy gauge group 
is given by
$SU(n_1) \times SU(n_2)\times U(1)$ or $SU(n)$, while
 gauge groups which are
products of more than two simple components (apart from $U(1)$) do not
seem to occur in this model. The values of $n_1$ and $n_2$ 
are determined dynamically.
Moreover, the existence of a magnetic flux in the vacua 
with non-simple gauge group
is very interesting in the context of fermions, since internal
fluxes naturally lead to chiral massless fermions. This will be
studied in detail elsewhere.

There is also an intriguing analogy between our toy model and string
theory, in the sense that as long as $a=0$,
there are a large number of possible
vacua (given by all possible partitions \eq{solution-general})
corresponding to compactifications, with no dynamical selection mechanism 
to choose one from the other. Remarkably this 
analog of the ``string vacuum problem'' is simply solved by 
adding a term to the action. 

Finally we should point out some potential problems or shortcomings of
our model. First, we have not yet fully established the existence of
the most interesting vacuum structure of type 3 such as in
\eq{vacuum-mod4} or \eq{vacuum-mod5}. This will be studied in a future
paper. Even a full analysis of the fluctuations and KK modes in the
scalar sector for vacuum of type 2 has not been done, but we expect no
surprises here; a numerical study is currently in progress. Finally,
the use of scalar Higgs fields $\phi_a$ without supersymmetry may seem
somewhat problematic due to the strong renormalization behavior of
scalar fields. This is in some sense consistent with the
interpretation as higher-dimensional gauge theory, which would be
non-renormalizable in the classical case. Moreover, 
a large value of the quadratically divergent term $\tilde b$ is
quite desirable here as explained in section \ref{sec:emergence},
and does not require particular fine-tuning.

\paragraph{Acknowledgements}

We are grateful for discussions with  C. Bachas, H. Grosse and B. Jurco.
This work was partially supported by the EPEAEK programme "Pythagoras"
and co-founded by the European Union (75\%) and the Hellenic state
(25\%).  The work of H.S. is supported by the FWF under project
P18657, and the work of P.A. is partially supported by
the EC contract MRTN-CT-2004-005104 and the italian 
MIUR contract PRIN-2005023102.   
H. S. in particular acknowledges an invitation to the
Technical University of Athens and to the Universita del Piemonte
Orientale, Alessandria. H.S, P.A. and G.Z. also acknowledge an
invitation to 
the Werner-Heisenberg-Institut Munich where some of this work has 
been carried out.

\section{Appendix}

\subsection{The fuzzy sphere}
\label{sec:fuzzysphere}

The fuzzy sphere \cite{Madore:1991bw} is a matrix approximation of the
usual sphere $S^2$. The algebra of functions on $S^2$ (which is
spanned by the spherical harmonics) is truncated at a given frequency
and thus becomes finite dimensional.  The algebra then becomes that of
$N\times N$ matrices. More precisely, the algebra of functions on the
ordinary sphere can be generated by the coordinates of $\R^3$ modulo
the relation $ \sum_{ {a}=1}^{3} {x}_{ {a}}{x}_{ {a}} =r^{2}$. The
fuzzy sphere $S^2_{N}$ is the non-commutative manifold whose
coordinate functions 
\be {x}_{ {a}} = r\,\frac {i}{\sqrt{C_2(N)}}\, {X}_{
{a}}, \qquad {x}_{ {a}}^\dagger = {x}_{ {a}}
\label{fuzzycoords}
\ee
are $N \times N$ hermitian matrices proportional to the generators of
the $N$-dimensional representation of $SU(2)$. They satisfy the
condition $\sum_{{a}=1}^{3} x_{{a}} x_{{a}} = r^2$ and the commutation
relations
\begin{equation}
[ X_{{a}}, X_{{b}} ] = \varepsilon_{abc}\, X_{{c}}~.
\end{equation}
For $N\rightarrow \infty$, one recovers the usual commutative sphere.
The best way to see this is to decompose the space of functions on
$S^2_{N}$ into irreps under the $SU(2)$ rotations, \bea S^2_{N} \cong
(N) \otimes (N) &=& (1) \oplus (3) \oplus ... \oplus (2N-1) \nn\\ &=&
\{Y^{0,0}\} \,\oplus \, ... \, \oplus\, \{Y^{(N-1),m}\}.
\label{fuzzyharmonics}
\eea This provides at the same time the definition of the fuzzy
spherical harmonics $Y^{lm}$, which we normalize as \be Tr_N
\left((Y^{lm})^\dagger Y^{l'm'}\right) = \delta^{l l'} \delta^{m m'}.
\ee Furthermore, there is a natural $SU(2)$ covariant differential
calculus on the fuzzy sphere. This calculus is three-dimensional, and
the derivations of a function $ f$ along $X_{{a}}$ are given by
$e_{{a}}({f})=[X_{{a}}, {f}]\,.\label{derivations}$ These are
essentially the angular momentum operators
\begin{equation}\label{LDA}
 J_a f = i{e}_{{a}} f = [i{X}_{{a}},f ],
\end{equation}
which satisfy the $SU(2)$ Lie algebra relation
 \begin{equation}
[J_a, J_b ] = i\varepsilon_{abc} J_c.
 \end{equation}
In the $N \rightarrow \infty$ limit the derivations $e_{{a}}$ become
$e_{{a}} = \varepsilon_{abc} x_{{b}}\partial_{{c}}$, and only in this
commutative limit the tangent space becomes two-dimensional. The
exterior derivative is given by
\begin{equation}
d f = [X_{{a}},f]\theta^{{a}}
\end{equation}
where $\theta^{{a}}$ are defined to be the one-forms dual to the
vector fields $e_{{a}}$,
$<e_{{a}},\theta^{{b}}>=\delta_{{a}}^{{b}}$. The space of one-forms is
generated by the $\theta^{{a}}$'s in the sense that any one-form can
be written as $\omega=\sum_{{a}=1}^3{\omega}_{{a}}\theta^{{a}}$.  The
differential geometry on the product space Minkowski times fuzzy
sphere, $M^{4} \times S^2_{N}$, is easily obtained from that on $M^4$
and on $S^2_N$. For example a one-form $A$ defined on $M^{4} \times
S^2_{N}$ is written as
\begin{equation}\label{oneform}
A= A_{\mu} dy^{\mu} + A_{{a}} \theta^{{a}}
\end{equation}
with $A_{\mu} =A_{\mu}(y^{\mu}, x_{{a}} )$ and $A_{{a}}
=A_{{a}}(y^{\mu}, x_{{a}} )$.

For further developments see
e.g. \cite{Balachandran:2005ew,Grosse:1995ar,Chu:2001xi}
and references therein.

\subsection{Gauge theory on the fuzzy sphere}
\label{sec:fuzzygaugetheory}

Here we briefly review the construction of YM gauge theory 
on $S^2_N$ as multi-matrix model
\cite{Steinacker:2003sd,Presnajder:2003ak,Carow-Watamura:1998jn}. 
Consider the action 
\bea
S &=& \frac{4\pi}{\cN}\, Tr \Big(a^2 (\phi_a\phi_a + C_2(N))^2 
+  \frac 1{\tilde g^2}\, F_{ab}^\dagger F_{ab}\Big)
\label{YM-FS}
\eea
where $\phi_a = - \phi_a^{\dagger}$ is an antihermitian $\cN \times
\cN$ matrix, 
and define\footnote{This can indeed be seen as components of 
the two-form $F = dA + AA$}
\be
F_{{a}{b}} = [\phi_{{a}}, \phi_{{b}}] 
-  \varepsilon_{abc} \phi_{{c}}\, .
\ee
This action is invariant under the $U(\cN)$ ``gauge'' symmetry acting as
$$
\phi_a \to U^{-1} \phi_a U. 
$$
A priori, we do not assume any underlying geometry, which arises
dynamically.
We claim that it describes $U(n)$ YM gauge theory on 
the fuzzy sphere $S^2_{N}$, assuming that
$\cN = N n$.

To see this, we first note that the action is positive definite,
with global minimum $S=0$ for the ``vacuum'' solution
\be
\phi_a = X_a^{(N)}\otimes \one_n
\ee
where $X_a \equiv X_a^{(N)}$ are the generators of the $N$- dimensional 
irrep of $SU(2)$. This is a first indication that the model 
``dynamically generates'' its own geometry, which is  
the fuzzy sphere $S^2_{N}$. In any case, 
it is natural to write a general field 
$\phi_a$ in the form
\be
\phi_a = X_a + A_a,
\label{gaugefield-S2N}
\ee
and to consider $A_a = \sum_\a A_{a,\a}(x) \,T_\a$ 
as functions $A_{a,\a}(x) = -A_{a,\a}(x)^\dagger$ 
on the fuzzy sphere $S^2_{N}$, taking value in $u(n)$ with generators $ T_\a$. 
The gauge transformation then takes the form
\bea
{A}_a &\to&  U^{-1} {A}_a U +  U^{-1} [X_a, U]\nn\\
 &=&  U^{-1} {A}_a U -i  U^{-1} J_{a} U,
\eea
which is the transformation 
rule of a  $U(n)$ gauge field. The field strength becomes
\bea
F_{ab} &=& [X_a,{ A}_b] -  [X_b,{A}_a] 
  + [{A}_a,{A}_b] - \varepsilon_{abc} {A}_c \nn\\
  &=&  -i J_{a} A_b  +i J_{b} A_a + [{A}_a,{A}_b] 
- \varepsilon_{abc} {A}_c.
\eea
This look like the 
field strength of a nonabelian $U(n)$ gauge field, 
with the caveat that we seem to have 3 degrees of freedom rather than 2.
To solve this puzzle, consider again the action, writing it in the form
\bea
S &=& \frac{4\pi}{\cN}\, Tr \Big(a^2 \varphi^2 
+ \frac 1{\tilde g^2}\, F_{ab}^\dagger F_{ab}\Big),
\label{YM-FS-2}
\eea
where we introduce the scalar field
\be
\varphi := \phi_a \phi_a + C_2(N) 
 = X_a {A}_{a} + {A}_{a} X_a + {A}_{a}{A}_{a}.
\ee
Since only configurations where $\varphi$ and $F_{ab}$ are small will
significantly contribute to the action, 
it follows that
\be
x_a {A}_a + {A}_a x_a = O(\frac{\varphi}N)
\label{A-constraint}
\ee
is small. This means that ${A}_{a}$ is tangential in the
(commutative) large $N$ limit, and 2 
tangential gauge degrees of freedom\footnote{to recover 
the familiar form of gauge theory, one needs to
rotate the components locally by $\frac{\pi} 2$ using the 
complex structure of $S^2$.
A more elegant way to establish the interpretation as YM action can be
given  using differential forms on $S^2_{N}$.}
survive. 
Equivalently, one can use the scalar field
$\phi = N \varphi$,
which would acquire a mass of order $N$ and decouple from the
theory.

We have thus established that the matrix model \eq{YM-FS} is indeed 
a fuzzy version of pure $U(n)$ YM theory on the sphere,
in the sense that it reduces to the commutative model in the large $N$
limit.
Without the term $(\phi_a \phi_a + C_2(N))^2$, the scalar field
corresponding to the radial component of $A_a$ no longer decouples 
and leads to a different model.

The main message to be remembered is the fact that the
matrix model \eq{YM-FS} without any further geometrical assumptions
dynamically generates the space $S^2_{N}$, and the fluctuations
turn out to be gauge fields governed by a $U(n)$ YM action.
Furthermore, the vacuum has no flat directions\footnote{the
  excitations 
turn out to be monopoles as
expected \cite{Steinacker:2003sd}, and fluxons 
similar as in \cite{Behr:2005wp}}, 
as we demonstrate explicitly in the following
section.

\subsection{Stability of the vacuum}
\label{sec:stability}

To establish stability of the vacua \eq{vacuum-mod1}, \eq{vacuum-mod2}
we should work out the spectrum of excitations around this 
solution and check whether there are flat or unstable modes. 
This is a formidable task in general, and we only consider the 
simplest case of the irreducible vacuum \eq{vacuum-mod1} 
for the case $\tilde b = C_2(N)$ and $d=0$ here. Once we 
have established that all fluctuation modes have strictly positive
eigenvalues, the same will hold in a neighborhood of this point
in the moduli space of couplings $(a,b,d,\tilde g, g_6)$.

An intuitive way to see this is by noting that the
potential $V(\phi_a)$ can be interpreted as YM gauge theory on 
$S^2_{N}$ with gauge group $U(n)$.
Since the sphere is compact, we expect that all fluctuations 
around the vacuum  $\phi_a = X_a^{(N)} \otimes \one_n$
have positive energy. We fix $n=1$ for simplicity.
Thus we write
\be
\phi_a =  X_a + A_a(x)
\ee 
where $A_a(x)$ is expanded into a suitable basis of
harmonics of $S^2_{N}$, which we should find.  It turns
out that a convenient way of doing this is to consider 
the antihermitian $2N \times 2N$ matrix \cite{Steinacker:2003sd}
\be 
\Phi = -\frac i2 + \phi_a \sigma_a = \Phi_{0} \, + A
\ee 
which satisfies 
\be
\Phi^2=  \phi_a \phi_a - \frac 14 
+ \frac i2 \varepsilon_{abc} F_{bc} \sigma_a.
\label{Phi2-F}
\ee
Thus $\Phi^2 = -\frac{N^2}4$ for $A=0$,
and in general we have 
\be
\tilde S_{YM} := Tr (\Phi^2 + \tilde b + \frac 14)^2
 = Tr \Big((\phi_a \phi_a  + \tilde b)^2 
+ F_{ab}^\dagger F_{ab} \Big).
\label{Stilde}
\ee
The following maps turn out to be useful:
\be
\cD(f) := i\{\Phi_0, f\}, 
\qquad \cJ(f) := [\Phi_0, f]
\ee
for any matrix $f$. The maps $\cD$ and $\cJ$ satisfy
\be
\cJ \cD = \cD \cJ =i[\Phi_0^2,.], \qquad \cD^2 - \cJ^2 = -2\{\Phi_0^2,.\},
\ee
which for the vacuum under consideration become
\be
\cJ \cD = \cD \cJ =0, \qquad \cD^2 - \cJ^2 = N^2 , \qquad
\cJ^3 = -N^2\, \cJ.
\ee
Note also that
\be
\cJ^2(f) = [\phi_a,[\phi_a,f]]  =: -\Delta f 
\ee
is the Laplacian, with eigenvalues
$\Delta f_l =  l(l+1) f_l$  (for the vacuum).

It turns out that the following is a natural basis of fluctuation modes:
\bea
\delta \Phi^{(1)} &=& A^{(1)}_a \sigma_a = \cD(f) - f, \nn \\
\delta \Phi^{(2)} &=&   A^{(2)}_a \sigma_a = \cJ^2(f') -\cJ^2(f')_0 
   = \cJ^2(f') +\Delta f'  \nn \\
\delta \Phi^{(g)} &=&  A^{(g)}_a \sigma_a = \cJ(f'') 
\label{basis} 
\eea 
for antihermitian $N \times N$ matrices $f,f',f''$, 
which will be expanded into 
orthonormal modes $f = \sum  f_{l,m}\, Y_{lm}$. 
Using orthogonality it is enough to 
consider these modes separately, i.e. $f = f_l = - f_l^\dagger$ 
with $Tr(f_l^\dagger f_l) =1$. 
One can show that
these modes form a complete set of fluctuations  
around $\Phi_0$ (for the vacuum). 
Here $A_{(g)}$ corresponds to gauge transformations, which we will omit
from now on.
Using
\be
Tr (f \cJ(g)) = - Tr (\cJ(f) g), 
\qquad Tr (f \cD(g)) = Tr f (\cD(f) g)
\ee
we can now compute the inner product matrix $Tr A^{(i)}A^{(j)}$:
\bea
Tr (A^{(1)}A^{(1)}) &=&  Tr (((N^2-1) f - \Delta(f)) g), \nn\\
Tr (A^{(1)}A^{(2)}) &=&   Tr(\Delta(f) g), \nn\\
Tr (A^{(2)}A^{(2)}) &=& Tr((N^2\Delta(f) -\Delta^2 f))g).
\eea
It is convenient to introduce
the matrix of normalizations for the modes $A^{(i)}$,
\bea
G_{ij}\equiv \Tr((A^{(i)})^\dagger A^{(j)})
= \left(\begin{array}{cc} (N^2-1)  - \Delta,
    & \Delta \\ 
  \Delta, & N^2\Delta -\Delta^2 
\end{array}\right)
\eea
which is positive definite except for the zero mode $l=0$
where $A^{(2)}$ is not defined.

We can now expand the action \eq{YM-FS} up to second order in these 
fluctuations. Since $F_{ab}=0$ and $(\phi_a \phi_a + \tilde b)=0$ 
for the vacuum, we have\footnote{Note that 
$\delta  Tr(\phi\cdot\phi)=0$ except for the zero mode $A^{(1)}_0$
with $l=0$ where
$\delta^{(1)} Tr(\phi\cdot\phi) \neq 0$,
as follows from \eq{phi2-fluct}. This mode corresponds 
to fluctuations of the radius, which will be discussed separately.}
\be
\delta^2 S_{YM}  = 
 Tr \Big(-\frac 1{\tilde g^2}\, \delta F_{ab} \delta F_{ab} 
 + a^2 \delta(\phi_a \phi_a)\delta(\phi_b \phi_b)\Big).
\ee
If $a^2\geq \frac 1{\tilde g^2}$, this can be written as
\bea
\delta^2 S_{YM}  &=& 
 Tr \Big(\frac 1{\tilde g^2}\, (-\delta F_{ab} \delta F_{ab} 
 + a^2 \delta(\phi_a \phi_a)\delta(\phi_a \phi_a))
  + (a^2-\frac 1{\tilde g^2})\delta(\phi_a \phi_a)\delta(\phi_a
  \phi_a) \Big) \nn\\
&=&  Tr\Big(\frac 1{\tilde g^2}\,\delta \Phi^2 \delta \Phi^2
+ (a^2-\frac 1{\tilde g^2})\delta(\phi_a \phi_a)\delta(\phi_a
  \phi_a)\Big) 
\eea 
and similarly for 
$a^2 < \frac 1{\tilde g^2}$. It is therefore enough to show that
\be
\delta^2 \tilde S_{YM}  = Tr(\delta \Phi^2 \delta \Phi^2)
= \Tr(-\delta^{(i)} F_{ab} \delta^{(j)} F_{ab} 
+ \delta^{(i)} (\phi\cdot\phi) \delta^{(j)} (\phi\cdot\phi) ) 
\ee
has a finite gap in the excitation spectrum. This spectrum can be computed 
efficiently as follows: note first
\bea
\delta^{(1)} \Phi^2 &=& -i\cD^2(f) +i \cD(f) 
                    = -i\cJ^2(f) +i \cD(f) -i N^2 f, \nn\\
\delta^{(2)} \Phi^2 &=& -i\cD(\Delta f), \nn\\
\delta^{(g)} \Phi^2 &=& -i\cD\cJ(f) = [\Phi_0^2,f] =0
\label{phi2-fluct}
\eea
for the vacuum. One then finds
\bea
Tr (\delta^{(1)} (\Phi^2) \delta^{(1)} (\Phi^2)) 
&=& - Tr(f)((-(N^2+1)\Delta  + (N^2 -1)N^2) g), \nn\\
Tr (\delta^{(1)} (\Phi^2) \delta^{(2)} (\Phi^2)) 
&=& -  Tr(f)  (\Delta^2)( g) , \nn\\
Tr (\delta^{(2)} (\Phi^2) \delta^{(2)} (\Phi^2)) 
&=& - Tr (g) (-\Delta^3 + N^2\Delta^2)g).
\eea
Noting that the antihermitian modes satisfy $Tr(f_l f_l) =-1$,
this gives
\be
\delta^2 \tilde S_{YM}  
= \left(\begin{array}{cc} 
-(N^2+1)\Delta + N^4-N^2, &  \Delta^2 \\ 
  \Delta^2 , & -\Delta^3  + N^2\Delta^2
\end{array}\right) = G T
\label{Stilde-fluct}
\ee
where the last equality defines $T$.
The eigenvalues of $T$ are found to be
$N^2$ and $\Delta$. These eigenvalues 
coincide\footnote{To see this, assume that we use an 
orthonormal basis  $A^o_{(i)}$ instead of the basis \eq{basis},
i.e. $A=b_1 A^o_{(1)}+b_2 A^o_{(2)}$. 
Then we can write $G=g^T g$ and $b_i=g_{ij} a_j$. Thus
\eq{Stilde-fluct} becomes $a^T\  G T\ a=b^T\ g\ T\ g^{-1} b$, and the 
eigenvalues of $ g\ T\ g^{-1}$ coincide with those of $T$, 
which therefore gives the masses.} with the spectrum
of the fluctuations of $\tilde S_{YM}$.
In particular, all modes with $l>0$ have positive mass.
The $l=0$ mode 
\be
A^{(1)}_0 = \cD(f_0) - f_0 = (2 i \Phi_0 -1) f_0
 = 2 i f_0\, \sigma_a \phi_a 
\ee
requires special treatment, and corresponds precisely to the 
fluctuations of the normalization $\a$, i.e. the 
radius of the sphere. We have shown explicitly in \eq{Higgs-potential} 
that this $\a = \a(y)$ has a positive mass.
Therefore we conclude that  all modes have positive mass, and 
there is no flat or unstable direction. This establishes the stability of
this vacuum.

The more general case  $\tilde b = C_2(N) + \epsilon$
with $\a \neq 1$ could be analyzed with the same methods,
which however will not be done in this paper.
For the reducible vacuum \eq{vacuum-mod2} or \eq{vacuum-mod3}
the analysis is more complicated, and will not be carried out here.

\bibliographystyle{diss}

\bibliography{mainbib}

\end{document}